\documentclass[a4paper]{jpconf}
\usepackage{graphicx}

\newcommand{\vp}{v_2(p_T)}
\newcommand{\etas}{\eta/s}

\begin{document}
\title{Collective Flows in a Transport Approach}

\author{S. Plumari$^{1,2}$, V. Baran$^3$, M. Di Toro$^{1,2}$ and V. Greco$^{1,2}$}

\address{Department of Physics and Astronomy, University of Catania, Via S. Sofia 64, I-95125 Catania}

\address{Laboratorio Nazionale del Sud, INFN-LNS, Via S. Sofia 63, I-95125 Catania}

\address{Physics Faculty, University of Bucharest and NIPNE-HH, Bucharest, Romania}

\ead{greco@lns.infn.it}

\begin{abstract}
We introduce a transport approach at fixed shear viscosity to entropy ratio $\etas$ to study the 
generation of collective flows in ultra-relativistic heavy-ion collisions. 
Transport theory supplies a covariant approach valid
also at large $\etas$ and at intermediate transverse momentum $p_T$, where deviations from equilibrium
is no longer negligible. 
Such an approach shows that at RHIC energies a temperature dependent $\etas$
enhances significantly the $v_4/v_2^2$ respect to the case of constant $\etas$. 
Furthermore if NJL chiral dynamics is self-consistently implemented we show that it does not modify
the relation between $v_2$ and $\etas$.

\end{abstract}

\section{Introduction}
The RHIC program at BNL has shown that the azimuthal asymmetry in
momentum space, namely the elliptic flow $v_2$, is the largest ever seen in HIC suggesting that an almost perfect fluid with a very small shear viscosity to entropy density ratio, $\etas$, has been created \cite{Adams:2005dq,Adcox}.
Nonetheless it has been found that dissipative effects cannot be neglected and even a small 
shear viscosity to entropy ratio $\etas$ produce sizeable effect increasing with the 
transverse momentum $p_T$ of the particles \cite{Romatschke:2007mq}. 
This has triggered a lot of activity in developing a relativistic theory of viscous hydrodynamics.
Viscous corrections to ideal hydrodynamics are indeed large and
a simple relativistic extension of first order Navier Stokes equations is affected by causality and stability
pathologies \cite{Huovinen:2008te}.
It is therefore necessary to go to second order gradient expansion,
and in particular the Israel-Stewart theory has been implemented to simulate the RHIC collisions providing an upper bound for $\eta/s\leq 0.4$
\cite{Song:2009rh}.
Such an approach, apart from the limitation to 2+1D simulations, has the more fundamental problem of a limited range of validity in $\etas$ and in the transverse momentum $p_T$.

On the contrary a relativistic transport approach has the advantage to be a 3+1D approach not based on a gradient expansion in viscosity that is valid also for large $\etas$ and for out of equilibrium momentum distribution allowing a reliable description also of the intermediate $p_T$ range where the important property of quark number scaling (QNS) of $v_2(p_T)$ has been observed \cite{Fries:2008hs}.
In this $p_T$ region viscous hydrodynamics breaks its validity because the relative deviation of the
equilibrium distribution function $\delta f/f_{eq}$ increases with $p_T^2$
becoming large already at $p_T \ge 3T \sim 1 GeV$ \cite{Song:2009rh}.
Furthermore Boltzmann-Vlasov transport theory distinguishes between
the short range interaction associated to collisions and long range interaction associated to the
field interaction, responsible for the change of the Equation of State (EoS) respect to
that of a free gas. This last feature allows to unify two main ingredients that are relevant for the
formation of collective flow. 
In ideal hydrodynamics the $\vp$ depends strongly on the EoS
namely on the sound velocity $c_s^2=dP/d\epsilon$ \cite{Bhalerao:2005mm}, 
while the mean free path $\lambda$ is assumed to
be vanishing. In the parton cascade approach the EoS is fixed to be the one of a free gas
$c_s^2=1/3=P/\epsilon$, on the other hand the mean free path $\lambda=1/\rho \sigma$ is finite.
In the first stage of RHIC the two different approaches were able to account for the large 
$\vp$ observed; in particular the parton cascade with large scattering cross section
predicted the saturation of $\vp$ vs $p_T$ \cite{Molnar:2001ux}.
Anyway once viscosity is finite both a finite $\lambda$ and the EoS are important for the 
generation of the momentum anisotropies.

The basic equation of transport for the (anti-) quark phase-space distribution function $f^\pm$
under a scalar field interaction can be written as:
\begin{eqnarray}
\label{VlasovNJL}
p^{\mu}\partial_{\mu} f^{\pm}(x,p)+M(x)\partial_{\mu} M(x) \partial_{p}^{\mu} f^{\pm}(x,p)=\mathcal{C}(x,p)
\end{eqnarray}
where the first term is related to the free streaming, the second term represents the effect
of a scalar field modifying the $\epsilon=3P$ relation (giving a finite interaction measure, see Fig.2 (left)) and the last term is the effect of the collisions directly associated to a finite $\lambda$ and therefore to a finite $\etas$. 

In the following we will show two results obtained with such an approach:
one related with the large $v_4/v_2^2$ ratio observed at RHIC and the other discussing the role
of the NJL chiral dynamics on the relation between $\vp$ and $\etas$.
We will show calculations for $Au+Au$ at 200 AGeV; the density distribution in coordinate
space is given by the standard Glauber model. The maximum initial temperature is $T=340$ MeV and the
initial time is $\tau_0=0.6$ fm/c as usually done also in hydrodynamical calculations.

\section{The large $v_4/v_2^2$ ratio}
We first consider the reduction of the transport approach to a cascade, i.e. neglect
the field interaction, included in the next section. In such a case the mass $M(x)=0$ in Eq.(1).
It has been shown that a fluid at finite $\etas$ can be studied within a transport approach  
renormalizing locally the scattering cross section according to $\etas \cdot \sigma_{tr}=<p>/15\rho $, 
for details see Ref.s \cite{Ferini:2008he,Greco:2008fs,Huovinen:2008te}. 
The system of course expands and cools and when the energy density is below $\epsilon=2 GeV/fm^3$
it starts the cross-over region and then the hadronic stage. During this phase we increase
the $\eta/s$ of the medium according to Fig. 1 (left) in order to have finally in the hadronic phase
an $\eta/s=8/4\pi\sim 0.65$ in agreement with several calculation of $\etas$. 

At RHIC it has been observed $v_4(p_T)/v_2^2(p_T)$ ratio of about 0.8-0.9 quite larger 
than first ideal hydrodynamical predictions of about 0.5.
More realistic hydrodynamical calculations show that  the ratio can be even
lower than 0.5 and viscosity does not strongly affect such a ratio but can bring it
up to about 0.6. Furthermore in Ref.\cite{Luzum:2010ae} it has been shown that such a ratio is almost independent on the initial eccentricity, on the viscosity, on the particle species. 
Therefore even if $v_4$ and $v_2$ can depend on several details such a ratio is a quite stable one 
and its large value should be considered as an oppurtunity to have some new information
on the system not yet spot.
\begin{figure}[ht]
\includegraphics[height=1.8in,width=2.5in]{etas-ene.eps}
\hspace{0.8cm}
\includegraphics[height=2.0in,width=2.5in]{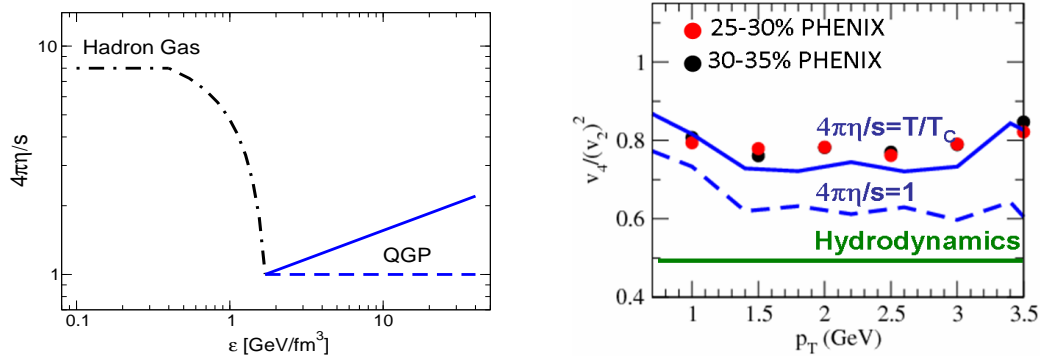}
\caption{Left: Dependence of $\etas$ on the energy density $\epsilon$ implemented in the 
transport code. Right: $v_4/v_2^2$ for $Au+Au$ collisions in the mid-rapidity
region $|y|<1$ at b= 8 fm; dashed line is the result for a constant $4\pi\etas=1$,
the solid line is the result for a temperature dependence of the $\etas$ like shown
by solid line on the left figure.}
\label{v4-v2}
\end{figure}
In Ref.\cite{Greco:2008fs} we have shown that indeed the formation time of $v_4$ is larger than the $v_2$
one and therefore it is more affected by dynamics of the later stage of the reaction.
Thanks to the difference in the formation time we argue that such a ratio is significantly
affected by the temperature dependence of the $\etas$.
Therefore we have performed two calculations one with a constant $4\pi\etas=1$ in the QGP phase (dashed line)
and the other, shown by
solid line in Fig. 1 (left), with a T or $\epsilon$ energy density dependence 
$4\pi \etas= T/T_0= (\epsilon/\epsilon_0)^{1/4}$ where $\epsilon_0= 1.7\, GeV/fm^3$ gives
the beginning of the cross over region. We remind that an increase of $\etas$ is expected
moving away from the phase transition region on general considerations \cite{Csernai:2006zz}.

In Fig.1 (left) we show the results of the two calculations for $Au+Au$ at 200 AGeV.
For a constant $\etas$ we find a value of about 0.6 while in the case 
of a temperature dependent $\etas$, it increases getting very close to experimental data,
$v_4/v_2^2 \sim $ 0.8.
The reason for the sensitivity of this ratio on the T dependence is that the elliptic flow
$v_2$ mainly develops earlier respect to $v_4$ and therefore on the average at larger
$\etas$ (for the solid line case), while $v_4$ develops more when the $\etas$ is smaller hence 
is less suppressed by dissipative effects.

\section{Impact of NJL Dynamics}
To start the investigation of the role of the field interaction we have implemented the NJL dynamics
that introduces the generation of a finite
mass associated to chiral symmetry breaking.
Therefore particles do not have a vanishing mass but a finite one according to
the mass gap equation of the NJL model:
\begin{eqnarray}
\label{NoneqMass2}
\frac{M(x)-m}{4\, g\, N_c}=M(x)\int \frac{d^3p}{(2\pi)^3} \frac{1- f^-(x,p)-f^+(x,p)}{E_p(x)}
\end{eqnarray}
that determines the local mass $M(x)$ at the space-time point $x$ in terms of the distribution functions $f^{\pm}(x,p)$. For a gas at finite mass $\epsilon - 3P$ is no longer vanishing as shown in Fig.2 (right)
for several NJL parametrization.

Eqs.(\ref{VlasovNJL}) and (\ref{NoneqMass2}) form a closed system of equations constituting
the Boltzmann-Vlasov equation associated to the NJL Lagrangian.
The space-time dependence of the mass $M_\alpha(r, t)=m-2g\big<\bar{\psi} \psi \big> $ influences the momenta of the particles because the finite gradient of the condensate generates a force which changes the momentum of a particle proportionally to $\vec{\nabla}_r \big<\bar{\psi} \psi \big>$.
The last is negative because the phase transition occurs earlier in the surface of the
expanding QGP fireball. Therefore
the phase transition which take place locally results in a negative contribution to the particle momenta
that makes the system more sticky respect to a free massless  gas.
To study a fluid at finite $\etas$ and NJL field interaction one has to extend the
simple relation $\sigma_{tr} \cdot \eta/s=<p>/15 \rho$
valid for a massless gas to the general case of massive relativistic particles and this has been
discussed in Ref.\cite{Plumari:2010ah}. 

\begin{figure}[ht]
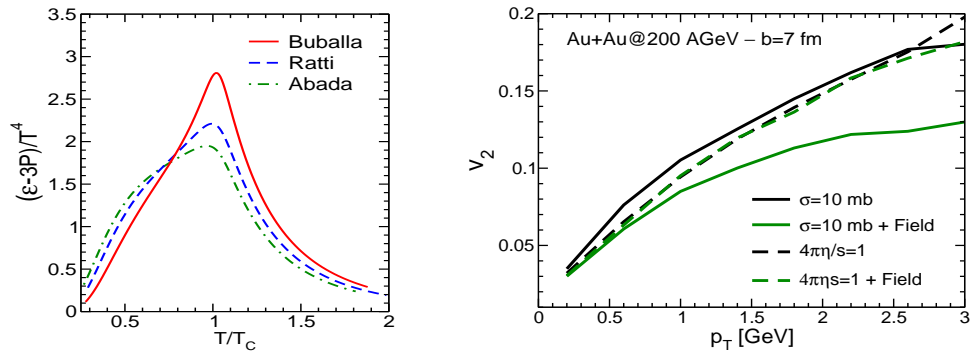

\vspace{0.4cm}
\includegraphics[height=1.8in,width=2.0in]{e3p-njl.eps}
\hspace{0.7cm}
\includegraphics[height=1.8in,width=2.6in]{v2-pT-b7-shear.eps}
\caption{Left: Interaction measure shown as a function of temperature T for three different
NJL parameter sets. Right: Average elliptic flow as a function of time for $Au+Au$ collisions in the mid-rapidity region $|y|<1$ at b= 7 fm; see text.}
\label{v2-pt}
\end{figure}

In Fig.\ref{v2-pt}, we show the elliptic flow as a function of $p_T$.
We find that the presence of an NJL-field that drives the chiral phase transition
suppress the $v_2(p_T)$ by about $20\%$ at $p_T > 1$ GeV.
This would imply the need of a parton scattering cross section $\sigma_{tr}$ even larger than
that estimated with the cascade model which was already quite larger than the pQCD estimates
\cite{Molnar:2001ux} in order to describe the data.
On the other hand the mean field modifies both the local entropy
density, reduced by the mass generation, and the shear viscosity that increases
respect to the massless case.
Considering that one of the main goal is to determine the $\eta/s$ of the QGP, we have investigated what
is the action of the mean field once the $\eta/s$ of the system is fixed to be the same
with and without the NJL field.
This of course implies different values of the cross section in the two cases.

The results for $4\pi\eta/s=1$ are shown by dashed lines in Fig. \ref{v2-pt} (right), the (black)
dashed line is the case with only the collision term (cascade)  while the (green) dashed line is the case with the field.
We can see that once the $\eta/s$ is fixed there is essentially no difference in the calculations
with and without a field dynamics included. This is a key result that shows that even in a microscopic approach that
distinguishes between the mean field and the collisional dynamics
the $v_2(p_T)$ is mainly driven by the $\eta/s$ of the fluid.
In other words we have found that in a microscopic approach the $\eta/s$ is the pertinent
parameter and the language of viscous hydrodynamics is appropriate. Of course this does not
mean that $v_2(p_T)$ in the transport theory is the same as in viscous hydrodynamics (especially
at $p_T > 1.5$ GeV), but
that the direct relation between $v_2(p_T)$ and $\eta/s$ is not modified by the NJL field dynamics. 
This confirms the validity of the studies pursued till
now even if they miss an explicit mean field dynamics and/or the chiral phase
transition.
 
\section*{Acknoweldgements}
This work for V. Baran is supported in part by the Romanian Ministry for Education and Research
under the CNCSIS contract PNII ID-946/2007.


\section*{References}
\begin{thebibliography}{9}

\bibitem{Adams:2005dq}
STAR, J.~Adams {\em et~al.},
\newblock Nucl. Phys. {\bf A757}, 102 (2005);

\bibitem{Adcox}
PHENIX, K.~Adcox {\em et~al.},
\newblock Nucl. Phys. {\bf A757}, 184 (2005).


\bibitem{Romatschke:2007mq}
P.~Romatschke and U.~Romatschke,
\newblock Phys. Rev. Lett. {\bf 99}, 172301 (2007).

\bibitem{Song:2009rh}
H.~Song and U.~W. Heinz,
\newblock (2009), 0909.1549.

\bibitem{Huovinen:2008te}
P.~Huovinen and D.~Molnar,
\newblock Phys. Rev. {\bf C79}, 014906 (2009).

\bibitem{Ferini:2008he}
G.~Ferini, M.~Colonna, M.~Di~Toro, and V.~Greco,
\newblock Phys. Lett. {\bf B670}, 325 (2009).

\bibitem{Greco:2008fs}
V.~Greco, M.~Colonna, M.~Di~Toro, and G.~Ferini,
\newblock Prog. Part. Nucl. Phys. {\bf 62}, 562 (2009).

\bibitem{Bhalerao:2005mm}
R.~S. Bhalerao, J.-P. Blaizot, N.~Borghini, and J.-Y. Ollitrault,
\newblock Phys. Lett. {\bf B627}, 49 (2005).

\bibitem{Molnar:2001ux}
D.~Molnar and M.~Gyulassy,
\newblock Nucl. Phys. {\bf A697}, 495 (2002).
 
\bibitem{Fries:2008hs}
R.~J. Fries, V.~Greco, and P.~Sorensen,
\newblock Ann. Rev. Nucl. Part. Sci. {\bf 58}, 177 (2008).

\bibitem{Luzum:2010ae}
  M.~Luzum, C.~Gombeaud and J.~Y.~Ollitrault,
  Phys.\ Rev.\  C {\bf 81} (2010) 054910
  
\bibitem{Csernai:2006zz}
  L.~P.~Csernai, J.~I.~Kapusta and L.~D.~McLerran,
  Phys.\ Rev.\ Lett.\  {\bf 97} (2006) 152303 

\bibitem{Plumari:2010ah}
  S.~Plumari, V.~Baran, M.~Di Toro, G.~Ferini and V.~Greco,
  Phys.\ Lett.\  B {\bf 689} (2010) 18

\end{thebibliography}
\end{document}